\documentclass{eas}
\usepackage{graphicx}
\usepackage{natbib}
\usepackage{caption}
\usepackage{subcaption}
\def\ho{\ifmmode {\rm HI} \else H{\small I} \fi}
\def\hh{\ifmmode {\rm H_2} \else H$_2$ \fi}
\def\no{\ifmmode {N_{\rm HI}} \else $N_{\rm HI}$ \fi}
\def\nt{\ifmmode {N_{\rm H_2}} \else $N_{\rm HI}$ \fi}
\def\so{\ifmmode {\Sigma_{\rm HI}} \else $\Sigma_{\rm HI}$ \fi}
\def\sht{\ifmmode {\Sigma_{\rm H_2}} \else $\Sigma_{\rm H_2}$ \fi}
\def\stot{\ifmmode {\Sigma_{\rm tot}} \else $\Sigma_{\rm tot}$ \fi}
\def\msun{\ifmmode {\rm M_{\odot}}\else $\rm M_{\odot}$\fi}
\def\mpc{\ifmmode {\rm M_{\odot} \ pc^{-2}} \else $\rm M_{\odot} \ pc^{-2}$ \fi}
\def\tra{\ifmmode {\rm HI-to-H_2}\else H{\small I}-to-H$_2$ \fi}
\def\aG{\ifmmode {\alpha G}\else $\alpha G$ \fi}
\def\sg{\ifmmode {\sigma_g}\else $\sigma_g$ \fi}
\def\sgt{\ifmmode {\sigma_{g-21}}\else $\sigma_{g-21}$ \fi}

\begin{document}

\title{H{\small I}-to-H$_2$ Transitions in the Perseus Molecular Cloud} 
\runningtitle{H{\small I}-to-H$_2$ Transitions in Perseus} 
\author{Shmuel Bialy}
\address{Raymond and Beverly Sackler School of Physics \& Astronomy, Tel Aviv University, Ramat Aviv 69978, Israel; \email{shmuelbi@mail.tau.ac.il}}
\author{Amiel Sternberg}
\sameaddress{1}
\author{Min-Young Lee}
\address{Laboratoire AIM, CEA/IRFU/Service d'Astrophyque, Bat 709, 91191 Gif-sur-Yvette, France}
\author{Franck Le Petit}
\address{LERMA, Observatoire de Paris, PSL Research University, CNRS, UMR8112, F-92190 Meudon, France}
\author{Evelyne Roueff}
\sameaddress{3}

\begin{abstract}

We apply the \citet[][hereafter S14]{Sternberg2014} theoretical model to analyze \ho and H$_2$ observations in the Perseus molecular cloud. We constrain the physical properties of the \ho shielding envelopes and the nature of the \tra transitions.
Our analysis \citep{Bialy2015} implies that in addition to cold neutral gas (CNM), less dense thermally-unstable gas (UNM) significantly contributes to the shielding of the H$_2$ cores in Perseus.

\end{abstract}
\maketitle
\section{Introduction}

Stars form in shielded molecular cores of giant molecular clouds. 
In Kennicutt-Schmidt relations, the star-formation rates correlate with molecular surface densities \citep[e.g.,~][]{Leroy2008, Genzel2013}.  
The conversion of atomic to molecular gas is also crucial for formation of other important molecular tracers and coolants such as CO, OH, and H$_2$O \citep[e.g.,~][]{Bialy2015a}.
Understanding the \tra transition is important for star-formation and galaxy evolution theories, and for interpreting observations of the ISM.

H$_2$ molecules are photodissociated by far-ultraviolet (FUV) radiation within the Lyman-Werner (LW) band (11.2 - 13.6 eV).
This occurs via a two-step process, in which a LW photon excites an electronic state, which in $\sim 12 \%$ of the cases decays to rovibrational continuum that leads to dissociation of the H$_2$ molecule. 
With increasing column density, the FUV radiation is absorbed, and the H$_2$ dissociation rate decreases.
Once the column density is large enough so that the local H$_2$ dissociation rate becomes equal to the H$_2$ formation rate, an \tra transition occurs.

What are the properties of \tra transitions in molecular clouds, and what are the properties of the (predominantly) \ho shielding columns? 

\section{Observations}
The Perseus cloud is located at a distance of $\sim 300$~pc, with an angular extent of $6^0 \times 3^0$, and a total mass of $\sim 10^4$~\msun \citep{Bally2008}.
Perseus consists of several dark and star-forming regions, which form low and intermediate mass stars (later than B1). Thus the FUV radiation in Perseus is probably dominated by external sources \citep[][hereafter L12]{Lee2012}. 

 L12 used 21 cm observations of the GALFA survey \citep{Peek2011}, together with IRIS infrared data \citep{MivilleDeschenes2005} and the $A_V$ image from the COMPLETE Survey \citep{Ridge2006}, to derive \ho and \hh surface densities (\so and $\sht$) towards B1, B1E, B5, IC348, and NGC1333, with a resolution of 0.4 pc.
We use the data presented by \citet{Lee2015}, for which the \ho columns were corrected for (up to 20 \%) 21 cm depth effects. 

\section{Theoretical Framework}

\setcounter{equation}{0}
\renewcommand\theequation{\arabic{equation}}

We apply the S14 theoretical model which assumes semi-infinite gas slabs irradiated by external FUV.
S14 derived an analytic formula for the total accumulated \ho surface column density, 
\begin{equation}
\label{eq: Sigma_HI}
\Sigma_{\rm HI} \ = \ 6.71 \ \Big(\frac{1.9}{\sgt}\Big) \ \ln \Big[ \frac{\aG}{3.2} \ + 1 \Big] \ \mpc , 
\end{equation}
Importantly, \so is independent of the total gas column \stot (or the cloud size), and is determined solely by the cloud physical parameters $\aG$ and $\sgt$.

Here, $\sgt$ is the dust cross section per hydrogen nucleus in units of $10^{-21}$~cm$^{2}$, and is typically $\approx 1.9$.
\aG is the (dimensionless) ratio of the H$_2$ {\it shielded} dissociation rate to H$_2$ formation rate.
Assuming H$_2$ formation on dust grains
\begin{equation}
\label{eq: aG}
\aG \ = \ 6 \ \Big( \frac{I_{UV}}{n/10 {\rm cm^{-3}}} \Big) \Big(\frac{w}{0.4} \Big) \ ,
\end{equation} where $I_{\rm UV}$ is the FUV intensity in units of the  \citet{Draine1978} field, $n$ is the volume density and $w$ is the fraction of LW photons that are absorbed in H$_2$-dust (see S14).
For multiphased gas, the CNM density and $I_{\rm UV}$ are proportional \citep{Wolfire2003}. In this case $(\aG)_{\rm CNM} \approx 3$. 
In our analysis however, we do not assume {\it a priori} CNM conditions \citep[as e.g.~][]{Krumholz2009}, but rather constrain $\aG$ directly using the observational data. 

\section{Results}

\begin{figure}
\includegraphics[width=1\textwidth]{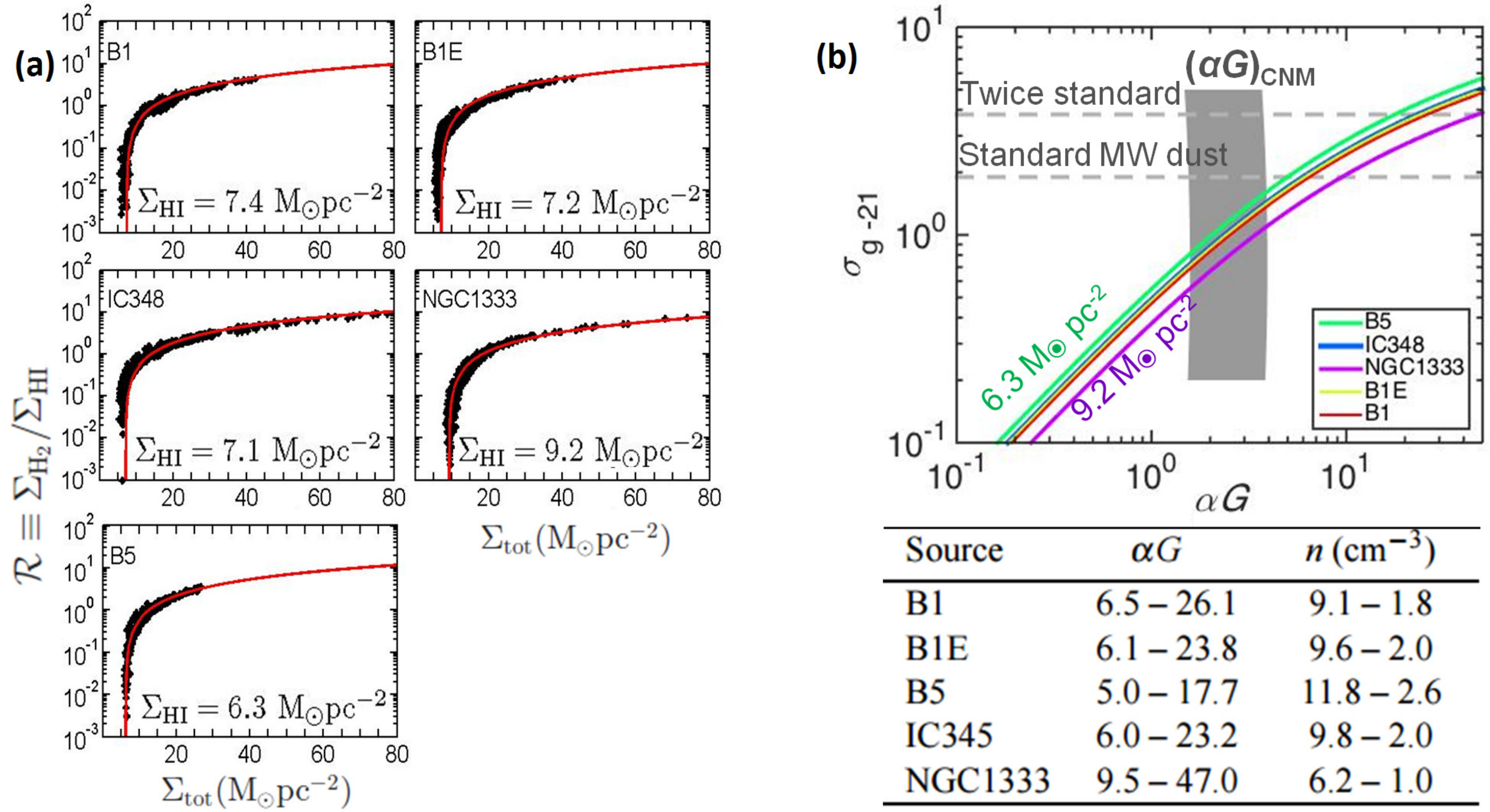}
\caption{(a) $R\equiv \so/\sht$ as a function of $\stot$, the black points are the observations and the red lines are fits to the model. The best fitted \ho columns are indicated. (b) The \so observed columns as contours in the $\aG$ -- $\sgt$ parameter space. The grey strip is the $(\aG)_{\rm CNM}$ typical range, and the two dashed horizontal lines are the typical $\sgt =1.9$ and twice typical (3.8) values. The $\aG$ values for this range and the corresponding densities (assuming $I_{\rm UV}=1$) are indicated.}
\end{figure}

In Fig.~1 (a) we test our theoretical prediction that $\so$ is independent of $\stot$ by fitting $\mathcal{R}\equiv\so/\sht$, for each of the five regions. The theory and observations are in excellent agreement. 
In Fig.~1 (b) we plot the \ho columns as contours in the $\aG$ -- $\sgt$ parameter space, using Equation (\ref{eq: Sigma_HI}).
L12 obtained an elevated $A_V/N_{\rm H}$ ratio in Perseus, so $\sgt$ probably lies within 1.9 to 3.8 (dashed lines).
For this realistic range in $\sgt$, $\aG$ spans from $\sim 5$ to
 $\sim 20$, a factor of 2 - 7 larger than $(\aG)_{\rm CNM}$ (grey strip). Therefore pure CNM shielding cannot explain the observed \ho columns in Perseus.

We use Equation (\ref{eq: aG}) to convert $\aG$ into volume densities $n$.
Assuming $I_{\rm UV} \approx 1$ \citep[][L12]{Tibbs2011}, we get $n \approx 2$ -- 10 cm$^{-3}$ for the \ho shielding layers in Perseus. These values are in-between the CNM and WNM densities, $n_{\rm CNM} \approx 100 n_{\rm WNM} \approx 22$~cm$^{-3}$ \citep{Wolfire2003}.

\section{Summary and Discussion}
We constrained the controlling parameter $\aG$ for the \ho envelopes in Perseus.
The $\aG$ and \ho volume densities are in-between the CNM and WNM values, suggesting that the \ho shielding layers are probably multiphased, where UNM (and perhaps some WNM) significantly contribute to the shielding of the H$_2$ cores.
An alternative explanation is that the observations of \so are contaminated by large amounts of \ho gas that does not participate in shielding. In this case $\so$ is effectively smaller, reducing the inferred $\aG$ and increasing $n$.
However, unrealistically large amounts of the \so must be removed (50-90\%) for all of the shielding gas to be CNM.
Therefore pure CNM shielding cannot explain the observed \ho columns in Perseus.

The situation in Perseus suggests that in addition to  CNM, less dense UNM is important in controlling the \tra transitions and Schmidt-Kennicutt thresholds in external galaxies.  
Full details of this work are in \citet{Bialy2015}.


\begin{thebibliography}{}

\bibitem[\protect\astroncite{Bally et~al.}{2008}]{Bally2008}
Bally, J., Walawender, J., Johnstone, D., et~al.: 2008,
\newblock {\em Handb. Star Form. Reg.} 4

\bibitem[\protect\astroncite{Bialy and Sternberg}{2015}]{Bialy2015a}
Bialy, S. and Sternberg, A.: 2015,
\newblock {\em MNRAS} {\bf 450(4)}, 4424

\bibitem[\protect\astroncite{Bialy et~al.}{2015}]{Bialy2015}
Bialy, S., Sternberg, A., Lee, M.-Y., et~al.: 2015,
\newblock {\em ApJ} {\bf 809(2)}, 122

\bibitem[\protect\astroncite{Draine}{1978}]{Draine1978}
Draine, B.~T.: 1978,
\newblock {\em ApJS} {\bf 36}, 595

\bibitem[\protect\astroncite{{Genzel} et~al.}{2013}]{Genzel2013}
{Genzel}, R., {Tacconi}, L.~J., {Kurk}, J., et~al.: 2013,
\newblock {\em ApJ} {\bf 773}, 68

\bibitem[\protect\astroncite{Krumholz et~al.}{2009}]{Krumholz2009}
Krumholz, M.~R., McKee, C.~F., and Tumlinson, J.: 2009,
\newblock {\em ApJ} {\bf 693(1)}, 216

\bibitem[\protect\astroncite{Lee et~al.}{2012}]{Lee2012}
Lee, M.-Y., Stanimirovi\'{c}, S., Douglas, K.~A., et~al.: 2012,
\newblock {\em ApJ} {\bf 748(2)}, 75

\bibitem[\protect\astroncite{{Lee} et~al.}{2015}]{Lee2015}
{Lee}, M.-Y., {Stanimirovic}, S., {Murray}, C.~E., et~al.: 2015,
\newblock {\em ApJ} {\bf 809(1)}, 56

\bibitem[\protect\astroncite{{Leroy} et~al.}{2008}]{Leroy2008}
{Leroy}, A.~K., {Walter}, F., {Brinks}, E., et~al.: 2008,
\newblock {\em AJ} {\bf 136}, 2782

\bibitem[\protect\astroncite{{Miville-Desch{\^e}nes} and
  {Lagache}}{2005}]{MivilleDeschenes2005}
{Miville-Desch{\^e}nes}, M.-A. and {Lagache}, G.: 2005,
\newblock {\em ApJS} {\bf 157(2)}, 302

\bibitem[\protect\astroncite{Peek et~al.}{2011}]{Peek2011}
Peek, J. E.~G., Heiles, C., Douglas, K.~A., et~al.: 2011,
\newblock {\em ApJS} {\bf 194(2)}, 20

\bibitem[\protect\astroncite{Ridge et~al.}{2006}]{Ridge2006}
Ridge, N.~A., {Di Francesco}, J., Kirk, H., et~al.: 2006,
\newblock {\em AJ} {\bf 131(6)}, 2921

\bibitem[\protect\astroncite{Sternberg et~al.}{2014}]{Sternberg2014}
Sternberg, A., Petit, F.~L., Roueff, E., and Bourlot, J.~L.: 2014,
\newblock {\em ApJS} {\bf 790}, 10S

\bibitem[\protect\astroncite{{Tibbs} et~al.}{2011}]{Tibbs2011}
{Tibbs}, C.~T., {Flagey}, N., {Paladini}, R., et~al.: 2011,
\newblock {\em MNRAS} {\bf 418}, 1889

\bibitem[\protect\astroncite{Wolfire et~al.}{2003}]{Wolfire2003}
Wolfire, M.~G., McKee, C.~F., Hollenbach, D., and Tielens, A. G. G.~M.: 2003,
\newblock {\em ApJ} {\bf 587(1)}, 278

\end{thebibliography}
\end{document}